\documentclass{article}
\usepackage{arxiv}

\usepackage{cite}
\usepackage{amsmath,amssymb,amsfonts}
\usepackage{algorithmic}
\usepackage{graphicx}
\usepackage{textcomp}

\usepackage{multirow}

\usepackage{verbatim} 
\usepackage{soul} 
\usepackage{wasysym} 

\usepackage[T1]{fontenc}
\usepackage[utf8]{inputenc}
\usepackage[emoticons]{emoji}

\title{Cross-Cultural Polarity and Emotion Detection Using Sentiment Analysis and Deep Learning - a Case Study on COVID-19}

\author{
Ali Shariq Imran\\
 Dept. of Computer Science, Norwegian University of Science \& Technology (NTNU), Norway.\\ ali.imran@ntnu.no\\
 \And
 Sher Muhammad Doudpota\\
 Dept. of Computer Science, Sukkur IBA University, Pakistan \\
 sher@iba-suk.edu.pk
\And
Zenun Kastrati \\
Dept. of Computer Science and Media Technology, Linnaeus University, Sweden,\\ zenun.kastrati@lnu.se\\
\And
Rakhi Bhatra \\
   Dept. of Computer Science, Sukkur IBA University, Pakistan \\
  rakhi.bhatra@iba-suk.edu.pk}

\begin{document}
\maketitle

\begin{abstract}
How different cultures react and respond given a crisis is predominant in a society's norms and political will to combat the situation. Often the decisions made are necessitated by events, social pressure, or the need of the hour, which may not represent the will of the nation. While some are pleased with it, others might show resentment. Coronavirus (COVID-19) brought a mix of similar emotions from the nations towards the decisions taken by their respective governments. Social media was bombarded with posts containing both positive and negative sentiments on the COVID-19, pandemic, lockdown, hashtags past couple of months. Despite geographically close, many neighboring countries reacted differently to one another. For instance, Denmark and Sweden, which share many similarities, stood poles apart on the decision taken by their respective governments. Yet, their nation's support was mostly unanimous, unlike the South Asian neighboring countries where people showed a lot of anxiety and resentment. This study tends to detect and analyze sentiment polarity and emotions demonstrated during the initial phase of the pandemic and the lockdown period employing natural language processing (NLP) and deep learning techniques on Twitter posts. Deep long short-term memory (LSTM) models used for estimating the sentiment polarity and emotions from extracted tweets have been trained to achieve state-of-the-art accuracy on the sentiment140 dataset. The use of emoticons showed a unique and novel way of validating the supervised deep learning models on tweets extracted from Twitter. 

\end{abstract}


\begin{figure}
  \centering
  \includegraphics[width = 16cm]{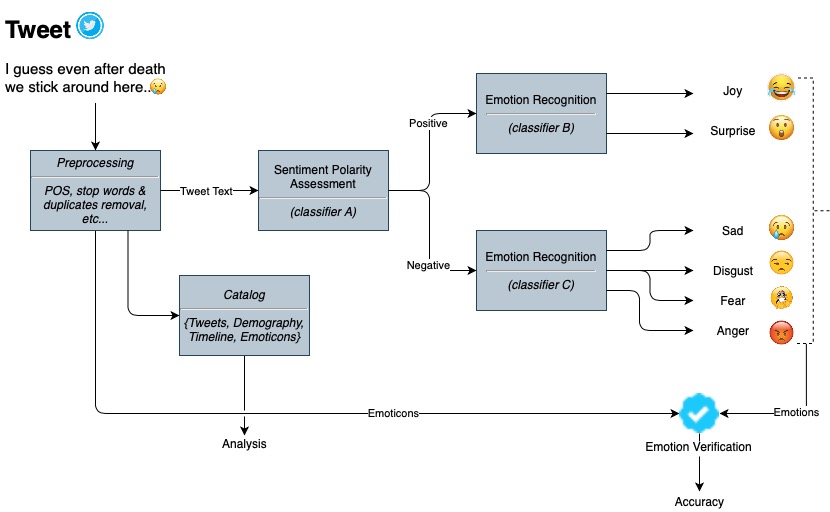}
  \caption{Abstract Model of the Proposed Tweets' Sentiment and Emotion Analyser.}
  \label{fig:model}
\end{figure}


\section{Introduction}
\label{sec:introduction}
The world is seeing a paradigm shift the way we conduct our daily activities amidst ongoing coronavirus (COVID-19) pandemic - be it online learning, the way we socialize, interact, conduct businesses or do shopping. Such global catastrophes have a direct effect on our social life; however, not all cultures react and respond in the same way given a crisis. Even under normal circumstances, research suggests that people across different cultures reason differently \cite{johnson2006there}. For instance, Nisbett in his book "The geography of thought: How Asians and Westerners think differently... and why" stated that the East Asians think on the basis of their experience dialectically and holistically, while Westerners think logically, abstractly, and analytically \cite{nisbett2004geography}. This cultural behavior and attitude are mostly governed by many factors, including the socio-economic situation of a country, faith and belief system, and lifestyle. In fact, the COVID-19 crisis showed greater cultural differences between countries that seem alike with respect to language, shared history and culture. For example, even though Denmark and Sweden are two neighboring countries that speak almost the same language and share a lot of culture and history, they stand at extreme ends of the spectrum when it comes to the way how they reacted to coronavirus \cite{Local:2020}. Denmark and Norway imposed more robust lockdown measures closing borders, schools, restaurants, and restricting gathering and social contact, while on the other side, Sweden has taken a relaxed approach to the corona outbreak keeping its schools, restaurants, and borders open. 

Social media platforms play an essential role during the extreme crisis as individuals use these communication channels to share ideas, opinions, and reactions with others to cope with and react to crises. Therefore, in this study, we will focus on exploring collective reactions to events expressed in social media. Particular emphasis will be given to analyzing people's reactions to global health-related events especially the COVID-19 pandemic expressed in Twitter's social media platform because of its widespread popularity and ease of access using the API. To this end, tweets collected from thousands of Twitter users communicated within four weeks after the corona crisis are analyzed to understand how different cultures were reacting and responding to coronavirus. Additionally, an extended version of publicly available tweets dataset was also used. A new model for sentiment and emotion analysis is proposed. The model takes advantage of natural language processing (NLP) and deep neural networks and comprises two main stages. The first stage involves sentiment polarity classifier that classifies tweets as positive and negative. The output of the first stage is then used as input to an emotion classifier that aims to assign a tweet to either one of positive emotions classes (joy and surprise) or one of the negative emotions classes (sad, disgust, fear, anger). Figure \ref{fig:model} shows the abstract model of proposed system of sentiment and emotion analysis on tweets' text.

\subsection{Study Objective \& Research Questions}
Our primary objective with this study is to understand how different cultures behave and react given a global crisis. The state of the questions addressed about the cultural differences as a techno-social system reveals potentialities in societal attitudinal, behavioral, and emotional predictions. 

In the present investigation, to examine those behavioral and emotional factors that describe how societies react under different circumstances, the general objective is to analyze the potential of utilizing NLP-based sentiment and emotional analysis techniques in finding answers to the following research questions (RQ). 

\begin{enumerate}
    \item RQ1: To what extent NLP can assist in understanding cultural behavior?
    \item RQ2: How reflective are the observations to the actual user sentiments analyzed from the tweets?
    \item RQ3: To what extent the sentiments are the same within and across the region? 
    \item RQ4: How are lockdowns and other measures seen by different countries/cultures? 
    
\end{enumerate}

\subsection{Contribution}
\label{sec:contribution}
The major contributions of this article are as following: 

\begin{itemize}
    \item A supervised deep learning sentiment detection model for Twitter feeds concerning the COVID-19 pandemic. 
    \item Proposed a multi-layer LSTM assessment model for classifying both sentiment polarity and emotions. 
    \item Achieved state-of-the-art accuracy on \textit{Sentiment140} polarity assessment dataset.
    \item Validation of the model for emotions expressed via emoticons. 
    \item Provide interesting insights into collective reactions on coronavirus outbreak on social media.
\end{itemize}{}

The rest of the article is organized as follows. Section \ref{sec:material_methods} presents the research design and study dimensions. Related work is presented in section \ref{sec:relaetdwork}. Data collection procedure and data preparation steps are described in section \ref{sec:dataset}, whereas, sentiment and emotion analysis model is presented in section \ref{sec:model-for-SAEA}. Section \ref{sec:results} entails the results followed by discussion and analysis in section \ref{sec:discussion}. Lastly, section \ref{sec:conclusion} concludes the paper with potential future research directions.

\section{Material \& Methods}
\label{sec:material_methods}

\subsection{Research Design}
The study is conducted using quantitative (experimental) research methodology on users' tweets posted post corona crisis. The investigation required collecting users' posts on Twitter from early February 2020 until the end of April 2020, when the first few cases were reported worldwide and in a respective country for ten to twelve weeks. The reason for using only the initial few weeks is that people usually get accustomed to the situation over time and an initial phase is enough to grasp the general/overall behavior of the masses towards a crisis and the policies adopted by respective governments. Several measurements have been taken in this study during data collection that requires cataloging for training deep learning models and for further analysis. These are discussed in the next subsection. 

\subsection{Study Dimensions}
Following dimensions are used to facilitate the interpretation of the results:

\begin{itemize}
    \item Demography-$(d)$: country / region under study. This study focuses on two neighbouring countries from South Asia, two from Nordic, and two from North America. 
    \item Timeline-$(t)$: the day from the initial reported cases in the country up to 4-12 weeks. 
    \item Culture-$(c)$: East (South-East Asia) vs. West (Nordic/America)
    \item Polarity-$(p)$: sentiment classified as either positive or negative. 
    \item Emotions-$(e)$: Feelings expressed as joy, surprise (astonished), sad, disgust, fear and anger. 
    \item Emoticons-$(et)$: emotions expressed through graphics for emotions listed above i.e., \emoji{1F600}, \emoji{1F632} \emoji{1F614}, \emoji{1F601}, \emoji{1F628},  \emoji{1F621}.   
    
\end{itemize}

\subsection{Tools \& Instrument}
Python scripts are used to query Tweepy Twitter API\footnote{https://www.tweepy.org} for fetching users' tweets and extracting feature set for cataloging. NLTK\footnote{https://www.nltk.org} is used to preprocess the retrieved tweets. NLP-based deep learning models are developed to predict sentiment polarity and users' emotions using Tensorflow and Keras as a back-end deep learning engine. \textit{Sentiment140} and \textit{Emotional Tweets} datasets are used to train classifier A and Classifier B/C respectively, as discussed in section \ref{sec:model-for-SAEA}. 
Visualization and LSTM model prediction as an instrument to analyze the results in addition to correlation are used. The results of sentiment and emotion recognition are validated through an innovative approach to exploiting emoticons extracted from the Tweets, which is a widely accepted feature of expressing one's feelings.  

\subsection{Deep Learning Models}
\label{sec:dlm}
Deep learning models for sentiment detection are employed in this study. A deep neural network (DNN) consists of an input, output, and a set of hidden layers with multiple nodes. The training process of a DNN consists of a pre-trainig and a fine-tuning steps. 

The pre-training step consists of weight initialization in an unsupervised manner via a generative deep belief networks (DBN) on the input data \cite{hinton2006fast}, followed by network training in a greedy way by taking two layers at a time as a restricted Boltzmann machine (RBM), given as:
\small
\begin{equation} 
\label{eq:RBM}
E(v,h)=-\sum_{k=1}^{K}\sum_{l=1}^{L}\frac{v_{k}}{\sigma_{k}}h_{l}w_{kl}-\sum_{k=1}^{K}\frac{(v_{k}-a_{k})^2}{2\sigma_{k}^{2}}-\sum_{l=1}^{L}h_{l}b_{l},
\end{equation}
\normalsize

\noindent where $\sigma_{k}$ is the standard deviation, $w_{kl}$ is the weight value connecting visible units $v_{k}$ and the hidden units $h_{l}$, $a_{k}$ and $b_{l}$ are the bias for visible and hidden units, respectively. The Equation \ref{eq:RBM} represents the energy function for the Gaussian-Bernoulli RMB. 

The hidden and visible units' joint-probability are defined as:
\begin{equation} 
\label{eq:JointProbability}
p(v,h)=\frac{e^{-E(v,h)}}{\sum\limits_{v,k} e^{-E(v,h)}} .
\end{equation}

Whereas, a contrastive divergence algorithm is used to estimate the trainable parameters by maximizing the expected log probability \cite{hinton2006fast}, given as:
\begin{equation} 
\label{eq:DivergenceAlg}
\theta=argmax_{\theta}E[log\sum\limits_{h}p(v,h)],
\end{equation}

\noindent where $\theta$ represents the weights, biases and standard deviation. 

The network parameters are adjusted in a supervised manner using back-propagation technique in the fine-tuning step. The back-propagation is an expression for the partial derivative $ \frac{\partial C}{\partial w}$ of the cost function $C$ with respect to any weight $w$ (or bias $b$) in the network. The quadratic cost function can be defined as:
\begin{equation} 
\label{eq:costFunction}
C = \frac{1}{2n}\sum_{x}^{} \left \| y(x) - \alpha^{L}(x) \right \|^{2},
\end{equation}

\noindent where $n$ is the total number of training examples, $x$
is the training samples, $y=y(x)$ is the corresponding desired output, $L$ denotes the number of layers in the network, and $\alpha^L=\alpha^L(x)$ is the vector of activations output from the network when $x$ is input.


The proposed sentiment assessment model employs LSTM, which is a variant of a recurrent neural network (RNN). LSTMs help preserve the error that can be back-propagated through time and layers. They allow RNN to learn continuously over many time steps by maintaining a constant error. RNN maintains memory which distinguish itself from the feedforward networks. LSTMs contain information outside the normal flow of the RNN in a gated cell. The process of carrying memory forward can be expressed mathematically as: 
\begin{equation} 
\label{eq:memory}
h_{t} = \phi (Wx_{t}+ Uh_{t-1}), 
\end{equation}

\noindent where $h_t$ is the hidden state at time $t$. $W$ is the weight matrix, and $U$ is the transition matrix. $\phi$ is the activation function. 




\section{Related Work}
\label{sec:relaetdwork}

\subsection{Reactions to Events in Social Media}
There is a large body of literature concerning people's reactions to events expressed in social media, which generally can be distinguished by the type of the event the response is related to and by the aim of the study \cite{Dunkel:2019}. Types of events cover natural disasters, health-related events, criminal and terrorist events, and protests, to name a few. Studies have been conducted for various purposes including examining the spreading pattern information on Twitter on Ebola \cite{Liang:2019} and on coronavirus outbreak \cite{Kaila:2020}, tracking and understanding public reaction during pandemics on twitter\cite{Szomszor:2011,Fu:2016}, investigating insights that Global Health can draw from social media \cite{Vorovchenko:2017}, conducting content and sentiment analysis of tweets \cite{Chew:2009}.

\subsection{Sentiment Polarity Assessment}
Sentiment analysis on Twitter data has been an area of wide interest for more than a decade. Researchers have performed sentiment polarity assessment on Twitter data for various application domains such as for donations and charity \cite{shelar2018sentiment}, students' feedback \cite{9110884}, on stocks \cite{das2018real, pagolu2016sentiment, batra2018integrating}, predicting elections \cite{budiharto2018prediction}, and understanding various other situations \cite{liao2017cnn}. Most approaches found in the literature have performed lexicon-based sentiment polarity detection via a standard NLP-pipeline (pre-processing steps) and POS tagging steps for SentiWordNet, MPQA, SenticNet or other lexicons. These approaches compute a score for finding polarity of the Tweet's text as the sum of the polarity conveyed by each of the micro-phrases $m$ which compose it \cite{musto2014comparison}, given as:
\begin{equation} 
\label{eq:DivergenceAlg}
Pol(m_i) = \sum_{j=1}^{k} \frac{score(term_j)\ast  w_{pos_(term_j)}}{\left | m_i \right |},
\end{equation}

\noindent where $w_{pos_(term_j)}$ is greater than 1 if $pos(term_j)$ = adverbs, verbs, adjectives, otherwise 1.

The abundance of literature on the subject cited led Kharde et al. \cite{kharde2016sentiment} and others \cite{zhang2018deep, giachanou2016like, desai2016techniques} to present a survey on conventional machine learning- / lexicon-based methods to deep learning-based technique respectively, to analyze tweets for polarity assessment, i.e., positive, negative, and neutral.

The authors in \cite{Xiang:2015} address the issue of spreading public concern about epidemics using Twitter data. A sentiment classification approach comprising two steps is used to measure people's concerns. The first step distinguishes personal tweets from the news, while the second step separates negative from non-negative tweets. To achieve this, two main types of methods were used: 1) an emotion-oriented, clue-based method to automatically generate training data, and 2) three different Machine Learning (ML) models to determine the one which gives the best accuracy.

Exploratory sentiment classification in the context of COVID-19 tweets is investigated in the study conducted by Samuel et al. \cite{Samuel:2020}. Two machine learning techniques, namely Na\"ive Bayes and Logistic Regression, are used to classifying positive and negative sentiment in tweets. Moreover, the performance of these two algorithms for sentiment classification is tested using two groups of data containing different lengths of tweets. The first group comprises shorter tweets with less than 77 characters, and the second one contains longer tweets with less than 120 characters. Na\"ive Bayes achieved an accuracy of 91.43\% for shorter tweets and 57.14\% for longer tweets, whereas, a worse performance is obtained by Logistic Regression, with an accuracy of 74.29\% for shorter tweets and 52\% for longer tweets, respectively. After the lockdown on the COVID-19 outbreak, Twitter sentiment classification of Indians is explored by the authors in \cite{Barkur:2020}. A total of 24,000 tweets collected from March $25^{th}$ to March $28^{th}$, $2020$ using the two prominent keywords: \#IndiaLockdown and \#IndiafightsCorona are used for analysis. The results revealed that even though there were negative sentiments expressed about the lockdown, tweets containing positive sentiments were quite present.

\subsection{Emotion Classification}

Hassan et al. \cite{hasan2014emotex} utilized the Circumplex model that characterizes affective experience along two dimensions: valence and arousal for detecting emotions in Twitter messages. The authors build the lexicon dictionary of emotions from emotional words from LIWC\footnote{http://www.liwc.net/} (Linguistic Inquiry \& Word Count). They extracted uni-grams, emoticons, negations and punctuation as features to train conventional machine learning classifiers in a supervised manner. They achieved an accuracy of 90\% on tweets.

The study conducted by Fung et al. \cite{Fung:2014} examines how people reacted to the Ebola outbreak on Twitter and Google. A random sample of tweets are examined, and the results showed that many people expressed negative emotions, anxiety, anger, which were higher than those expressed for influenza. The findings also suggested that Twitter can provide valuable information on people's anxiety, anger, or negative emotions, which could be used by public authorities and health practitioners to provide relevant and accurate information related to the outbreak.

The authors in \cite{do2016analyzing} investigate people's emotional response during the Middle East Respiratory Syndrome (MERS) outbreak in South Korea. They used eight emotions to analyze people's responses. Their findings revealed that 80\% of the tweets were neutral, while anger and fear dominated the tweet concerning the disease. Moreover, the anger increased over time, mostly blaming the Korean government while there was a decline in fear and sadness responses over time. This observation, as per the authors, was understandable as the government was taking strict actions to prevent the infection, and the number of new MERS cases decreased as time went by. The important finding was that the surprise, disgust, and happiness were more or less constant. A similar study is conducted by the researchers in \cite{Kaila:2020}. The study focuses on emotional reactions during the COVID-19 outbreak by exploring the tweets. A random sample of 18,000 tweets is examined for positive and negative sentiment along with eight emotions, including anger, anticipation, disgust, fear, joy, sadness, surprise, trust. The findings showed that there exists an almost equal number of positive and negative sentiments, as most of the tweets contained both panic and comforting words. The fear among the people was the number one emotion that dominated the tweets, followed by the trust of the authorities. Also, emotions such as sadness and anger of people were prevalent.




\begin{figure}
  \centering
  \includegraphics[width = 16cm]{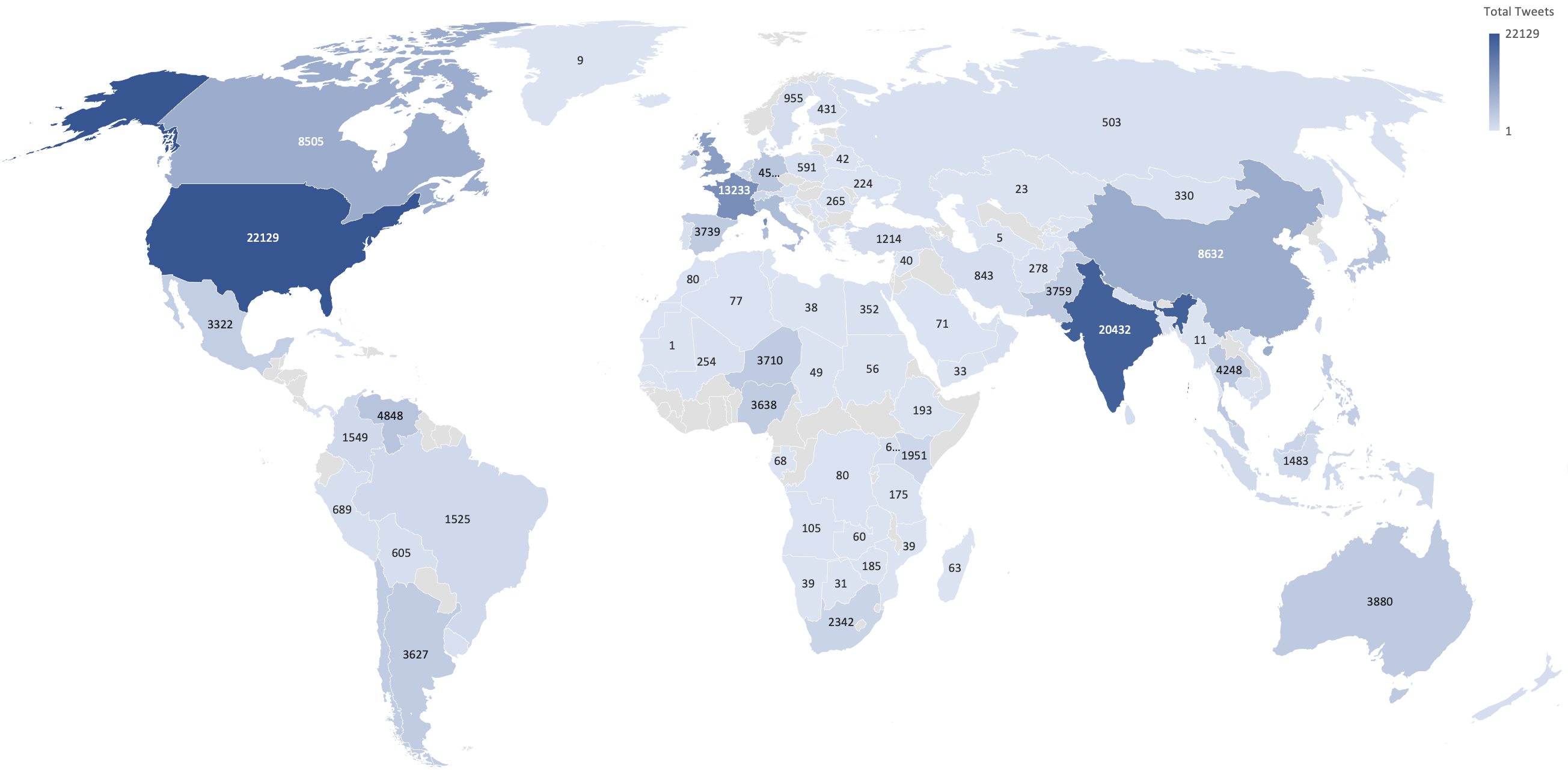}
  \caption{Total No. of Tweets per Country for the Period $3^{rd}$ to $29^{th}$ Feb. 2020 for Trending Hashtags \#.}
  \label{fig:world-map}
\end{figure}

\section{Dataset}
\label{sec:dataset}
We used two tweets' datasets in this study to detect sentiment polarity and emotion recognition. Trending hashtag \# data explained in subsection \ref{sec:thd} that we collected ourselves and the Kaggle dataset presented in subsection \ref{sec:k_dataset}. We additionally used the Sentiment140 \cite{go2009twitter} and Emotional Tweets dataset \cite{MohammadB17wassa} to train our proposed deep learning models. The reason for using these two particular datasets for training the model is: (i) the availability of manually labeled state-of-the-art dataset and (ii) the lack of labeled tweets extracted from Twitter.  The focus of our study is six neighboring  countries from three continents having similar cultures and circumstances. These include Pakistan, India, Norway, Sweden, USA, and Canada. We specifically opted for these six countries for cross-cultural analysis due size, approach adopted by respective governments, popularity and cultural similarity. 

\subsection{Trending Hashtag Data}
\label{sec:thd}
The study employs retrieving and collecting trending hashtag \# tweets ourselves due to the lack of publicly available datasets for the initial period of COVID-19 outbreak. For instance, \#lockdown was trending across the globe during February 2020; \#StayHome was trending in Sweden, while COVID-19 was trending throughout the period February - April 2020. Figure \ref{fig:world-map} shows the total number of tweets per country for trending hashtags \# between $3^{rd}$ February to $29^{th}$ February 2020. We only retrieved the trending hashtag \# tweets from across six countries mentioned earlier for the initial phase of the pandemic for this study.

\subsubsection{Data collection procedure}
\label{sec:dcp}
A standard Twitter search API, known as Tweepy, is used to fetch users' tweets. Multiple queries are executed via Tweepy containing trending keywords \#Coronavirus, \#COVID\_19, \#COVID19, \#COVID19Pandamic, \#Lockdown, \#StayHomeSaveLives, and \#StayHome for the period $T_p = \{S_d, E_d\}$, where $S_d$ is the starting date, i.e. when the first case of the corona patient is reported in a given country/region and $E_d$ is the end date. The keywords are chosen based upon the trending keywords during $T_p$. Only tweets in English for a given region are cataloged for further processing containing Tweet ID, text, user name, time, and location.  

\subsubsection{Data preparation}
\label{sec:dp}
PRISMA\footnote{http://www.prisma-statement.org} approach is adopted in this study to query COVID-19 related tweets and to filter out the irrelevant ones. Following pre-processing steps are applied to clean the retrieved tweets:

\begin{enumerate}
    \item Removal of mentions and colons from tweet text. 
    \item Replacement of consecutive non-ASCII characters with space.
    \item Tokenization of tweets.
    \item Removal of stop-words and punctuation via NLTK library. 
    \item Tokens are appended to obtain cleaned tweets. 
    \item Extraction of emoticons from tweets. 
\end{enumerate}

The following items are cataloged for each tweet: Tweet ID, Time, Original Text, Cleaned Text, Polarity, Subjectivity, User Name, User Location and Emoticons. A total of 27,357 tweets were extracted after pre-processing and filtering, as depicted in Table \ref{tab:NoOfTweets}.


\begin{table}[h]
 \centering
\begin{tabular}{ |c|l|c|c| } 
 \hline
 Sr.\#  & Country & Trending Hashtag \# Dataset & Kaggle Dataset  \\
 \hline
 
 \hline
 1 & Pakistan & 2501  & 9869 \\ 
 \hline
 2 & India & 8455 & 70392\\ 
\hline
 3 & Norway & 168 & 476 \\ 
 \hline
 4 & Sweden & 571 & 816\\ 
 \hline
 5 & Canada & 5367 & 42127 \\ 
 \hline
 6 & USA & 10295 & 336606 \\
 \hline 
 
 \hline
  & Total & 27357 & 460286 \\
 \hline
 \end{tabular}
 
 \caption{No of tweets per Country for Trending Hashtag \# Dataset After Filtering and Kaggle Dataset.}
\label{tab:NoOfTweets}
\end{table}

\subsection{Kaggle Dataset}
\label{sec:k_dataset}

We further went on to include Tweets for the period of March to April 2020 from the publically available dataset since after data-preparation, we were left with a small number of tweets from Nordic countries. Table \ref{tab:NoOfTweets} shows the number of tweets per country under consideration for the Kaggle dataset\footnote{https://www.kaggle.com/smid80/coronavirus-covid19-tweets} from $12^{th}$ of March to $30^{th}$ April 2020. The total number of tweets is 460,286, out of which USA tweets contribute 73\%. The hashtags \# applied to retrieve Kaggle dataset tweets include \#coronavirus, \#coronavirusoutbreak, \#coronavirusPandemic, \#covid19, \#covid\_19. From $17^{th}$ March till the end of the April two more hashtags were included, i.e., \# epitwitter, \#ihavecorona.


\subsection{Sentiment140 Dataset}
\label{sec:sentiment140_dataset}
We used the Sentiment140 dataset from Stanford \cite{go2009twitter} for training our sentiment polarity assessment classifier - A, presented in section \ref{sec:classifierA}. This dataset contains an overwhelming number of positive and negative tweets. Each category contains 0.8 million tweets, a staggering number of a total of 1.6 million tweets. We particularly opted for this dataset to train our deep learning models in a supervised manner due to the unavailability of the labeled tweets related to COVID-19. 

\subsection{Emotional Tweets Dataset}
\label{sec:etd_dataset}

Emotional Tweets dataset is utilized in this study to train classifier B and classifier C for emotions recognition, described in \ref{sec:er-C-B} and \ref{sec:er-C-C}, respectively. The tagging process of this dataset is reported by Saif et al. in \cite{MohammadB17wassa}. The dataset contains six classes as summarized in Table \ref{table_emotion_dataset}. The first two labels, joy and surprise, are positive emotions, whereas the remaining four, sadness, fear, anger, and disgust, are negative emotions. The dataset comprises of 21,051 total number of labeled tweets. 

\begin{table}[h]
 \centering
\begin{tabular}{ |c|l|c|c| } 
 \hline
  Sr. \# & Class Label & Number of Instances & Sentiment Polarity \\
 \hline
 
 \hline

 1 & Joy  & 8240 & Positive \\ 
 \hline
 2 & Surprise & 3849& Positive \\ 
\hline
 3 & Sad & 3830 & Negative \\ 
 \hline
 4 & Fear & 2816 & Negative\\ 
 \hline
 5 & Anger & 1555 & Negative \\ 
 \hline
 6 & Disgust & 761 & Negative \\
 \hline
 \end{tabular}
 
 \caption{Emotional Tweet Dataset Containing Six Class Labels for Positive and Negative Sentiment Polarity.}
\label{table_emotion_dataset}
\end{table}


\section{Model for Sentiment and Emotion Analysis}
\label{sec:model-for-SAEA}
Literature suggests many attempts of tweets' sentiment analysis, but very few attempts of emotions' classification. Sentiment analysis on tweets refers to the classification of an input tweet text into sentiment polarities, including positive, negative and neutral, whereas emotions’ classification refers to classifying tweet text in emotions’ label including joy, surprise, sadness, fear, anger and disgust.

Sentiment polarity certainly conveys meaningful information about the subject of the text; however, the emotion classification is the next level. It suggests if the sentiment about the subject is negative, then to what extent it is negative – being negative with anger is a  different state of mind than being negative and disgusted. Therefore, it is important to extend the task of sentiment polarity classification to the next level and identify emotion in negative and positive sentiment polarities. The rest of this section explains the working of each of the components in the abstract model, depicted in Figure \ref{fig:model}. All the models and Jupyter Notebooks developed for this paper are available on paper's GitHub repository\footnote{https://github.com/sherkhalil/COVID19}.  

\begin{table*}[t]
  \centering
  \begin{tabular}{|l | p{2.5cm}|p{6cm}|c|c|} 
  \hline
    Sr. \# & Model Name & Model Configuration/ Parameters & Training Accuracy & Validation Accuracy \\
    
    \hline
    
    \hline
    1 & DNN (Baseline)  & Embedding Layer with 300 Dimension, GlobalMaxPooling Layers with 128, 64, 32 with ReLU, Dense 2 with Sigmoid & 95\% & 81\% \\
    \hline
    2 & LSTM + FastText  & Embedding Layer with wiki-news-300d-1M.vec, LSTM(32) ReLU, Dropout and Recurrent Dropout = 0.2, Dense(2) with Sigmoid & 94\% & 82\% \\
    \hline
    3 & LSTM + GloVe  & Embedding Layer with glove.6B.300d.txt, LSTM(32) ReLU, Dropout and Recurrent Dropout = 0.2, Dense(2) with Sigmoid & 96\% & 82\% \\
    \hline
    4 & LSTM + GloVe Twitter  & Embedding Layer with glove.twitter.27B.200d.txt, LSTM(32) ReLU, Dropout and Recurrent Dropout = 0.2, Dense(2) with Sigmoid & 92\% & 83\% \\
    \hline
    5 & LSTM without pretrained embedding & Embedding Layer LSTM (32) with ReLU and Dense Layers 2 with Sigmoid & 96\% & 83\% \\
    \hline
  \end{tabular}
  \caption{Training-validation Accuracy on Sentiment140 Dataset for Five Proposed Deep Learning Models.}
  \label{models_training_table}
\end{table*}

\subsection{Sentiment Assessment -- Classifier A}
\label{sec:classifierA}
The first stage classifier in our model classifies an input tweet text in either positive or negative polarity. For this, we employed the Sentiment140 dataset explained in section \ref{sec:sentiment140_dataset} -- the most popular dataset for such polarity classification tasks. 
For developing our first stage model, we padded each input tweet to ensure a uniform size of 280 characters, which is standard tweet maximum size.

\begin{figure}
  \centering
  \includegraphics[width = 9.5cm]{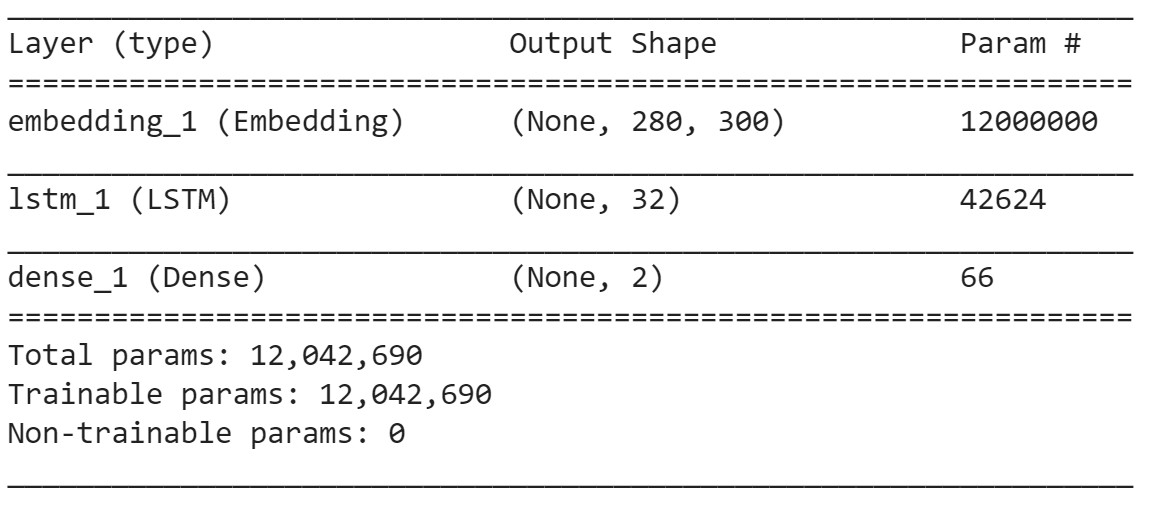}
  \caption{LSTM + FastText Model Summary.}
  \label{model_summary}
\end{figure}

To establish a baseline model, a simple deep neural network based on an embedding layer, max-pooling layer, and three dense layers of 128, 64, and 32 outputs were developed. The last layer uses sigmoid as an activation function, as it performs better in binary classification, whereas all intermediate layers use ReLU as an activation function. This baseline model splits 1.6 million tweets in training and test sets with 10\% tweets (160,000 tweets) spared for testing the model. The remaining 90\% tweets were further divided into a 90/10 ratio for training and model validation, respectively. 
The model training and validation was set to ten epochs; however, the model over fits immediately after two epochs, therefore, it was retrained on two epochs to avoid overfitting. The training and validation accuracy on the baseline models was 96\% and 81\%, respectively. Table \ref{models_training_table} summarizes training and validation accuracy for each of the five proposed models along with model structures. Figure \ref{model_summary} shows structure of the best performing model i.e. LSTM with FastText model.


\begin{table}[t]
 \centering
\begin{tabular}{ |c|p{3.7cm}|c|c| } 
 \hline
 
 \hline
 
  Model \# & Model Name & F1 Score & Accuracy \\
 \hline
 1 & DNN (Baseline) & 79.0\%  & 78.4\% \\ 
 \hline
  2 & LSTM + FastText & \textbf{82.4\%} & \textbf{82.4\%} \\ 
 \hline
  3 & LSTM + GloVe & 81.5\% & 81.4\% \\ 
 \hline
 4 & LSTM + GloVe Twitter & 80.4\% & 80.4\% \\ 
\hline

 5 & LSTM + w/o Pretrained Embed. & 81.6\% & 81.4\% \\ 
 \hline
 6 & CONV Based on \cite{caisentiment} & 81.7\%& 81.1\% \\
 \hline
 \end{tabular}
 \caption{F1 and accuracy scores of Six Deep Learning Models}
\label{table_model_accuracy}
\end{table}

Table \ref{table_model_accuracy} shows the F1 and the accuracy scores on test set -- 10\% of the dataset comprising of 160,000 tweets equally divided into positive and negative polarities. The table also presents the previously best-reported accuracy and F1 score on the dataset, as reported in \cite{caisentiment}. The model proposed in this paper based on FastText outperforms all other models, including previously best-reported accuracy. Therefore, we choose this model as our first stage classifier to classify tweets in positive and negative polarities.

\begin{table}[t]
 \centering
\begin{tabular}{ |l|p{3.7cm}|c|c| } 
 \hline
  Model \# & Model Name & F1 Score & Accuracy \\
 \hline
 
 \hline
 1 & DNN (Baseline) & 62.7\%	& 78.4\%\\ 
 \hline
 2 & LSTM + FastText & 67.5\% &	80.8\% \\ 
 \hline
 3 & LSTM + GloVe & 69.0\%  &	80.3\% \\ 
 \hline
 4 & LSTM + GloVe Twitter & \textbf{69.9\%}	 & \textbf{81.9\%} \\ 
\hline
 5 & LSTM + w/o Pretrained Embed. & 68.4\% &	79.8\% \\ 
 \hline

 \end{tabular}
 \caption{F1 and Accuracy Scores of Five Proposed Models on Positive Emotions (Joy and Surprise).}
\label{table_model_accuracy_joy_surprise}
\end{table}

\begin{table}[t]
 \centering
\begin{tabular}{ |c|p{3.7cm}|c|c| } 
 \hline
 Model \# & Model Name & F1 Score & Accuracy \\
 \hline
 
 \hline
 1 & DNN (Baseline) & 59.0\% &	64.5\%\\ 
 \hline
 2 & LSTM + FastText & 62.1\% &  66.0\%\\ 
 \hline
 3 & LSTM + GloVe & 65.8\% & 67.7\% \\ 
 \hline
 4 & LSTM + GloVe Twitter &\textbf{ 69.2\%}	& \textbf{ 69.9\%} \\ 
\hline
 5 & LSTM + w/o Pretrained Embed. & 62.1\%	& 66.0\% \\ 
 \hline
  \end{tabular}
 \caption{F1 and Accuracy Scores of Five Models on Negative Emotions (Sad, Anger, Fear)}
\label{table_model_accuracy_sadness_anger_fear}
\end{table}

\subsection{Emotion Recognition -- Classifier B}
\label{sec:er-C-B}
Once the polarity from Classifier A is positive, the next step is to identify positive emotions in the tweet. In order to extract tweet emotions, we use the Emotional Tweets dataset presented in section \ref{sec:etd_dataset}. 
If the label from first stage Classifier A is positive, the text is applied to classifier A to determine exact positive emotions -- joy or surprise.  

In order to extract positive emotions from the positive tweets, the negative emotions' labels were removed at Classifier B, leaving only two positive labels -- joy and surprise. Repeating the same experiments as in Classifier A, the performance of five models was tested for this classification task.  
The test accuracy for each of these models is reported in Table \ref{table_model_accuracy_joy_surprise}. The model based on Glove.twitter.27B.300d pre-trained embedding with LSTM outperforms the other four models; therefore, we use LSTM with GloVe embedding at this stage.

\begin{table}[!h]
 \centering
\begin{tabular}{| l | l | c | c | l |}
\hline
Sr                 & Emtotions                 & Emoticons & Unicode & Description                                         \\ \hline
\multirow{8}{*}{1} & \multirow{8}{*}{Joy}      & \emoji{1F600}     & 1F600   & grinning face                                       \\ \cline{3-5} 
                   &                           & \emoji{1F602}    & 1F602   & face with tears of joy                              \\ \cline{3-5} 
                   &                           & \emoji{1F603}     & 1F603   & smiling face with open mouth                        \\ \cline{3-5} 
                   &                           & \emoji{1F604}     & 1F604   & smiling face with open mouth and open eyes          \\ \cline{3-5} 
                   &                           & \emoji{1F605}     & 1F605   & smiling face with open mouth and cold sweat         \\ \cline{3-5} 
                   &                           &\emoji{1F606}     & 1F606   & smiling face with open mouth and tighly-closed eyes \\ \cline{3-5} 
                   &                           & \emoji{1F60A}     & 1F60A   & smiling face with smiling eyes                      \\ \cline{3-5} 
                   &                           & \emoji{1F60D}     & 1F60D   & smiling face with heart-shaped eyes                 \\ \hline
\multirow{3}{*}{2} & \multirow{3}{*}{Surprise} & \emoji{1F632}     & 1F632   & astonished face                                     \\ \cline{3-5} 
                   &                           & \emoji{1F62E}    & 1F62E   & face with open mouth                                \\ \cline{3-5} 
                   &                           & \emoji{1F62F}     & 1F62F   & hushed face                               \\ \hline
\multirow{6}{*}{3} & \multirow{6}{*}{Sad}      & \emoji{1F613}     & 1F613   & face with cold sweat                                \\ \cline{3-5} 
                   &                           & \emoji{1F614}     & 1F614   & pensive face                                        \\ \cline{3-5} 
                   &                           & \emoji{1F61E}     & 1F61E   & dissappointed face                                  \\ \cline{3-5} 
                   &                           & \emoji{1F622}     & 1F622   & crying face                                         \\ \cline{3-5} 
                   &                           & \emoji{1F62D}     & 1F62D   & loudly crying face                                  \\ \cline{3-5} 
                   &                           & \emoji{1F623}     & 1F623   & persevering face                                    \\ \hline
\multirow{3}{*}{4} & \multirow{3}{*}{Anger}    & \emoji{1F620}     & 1F620   & angry face                                          \\ \cline{3-5} 
                   &                           & \emoji{1F621}     & 1F621   & pounting face                                       \\ \cline{3-5} 
                   &                           & \emoji{1F624}     & 1F624   & face with look of triumph                           \\ \hline
\multirow{2}{*}{5} & \multirow{2}{*}{Fear}     & \emoji{1F628}     & 1F628   & fearful face                                        \\ \cline{3-5} 
                   &                           & \emoji{1F632}     & 1F632   & face screaming in fear                              \\ \hline
6                  & Disgust                   & \emoji{1F62C}     & 1F62C   & grimacing face                                      \\ \hline
\end{tabular}
 \caption{Grouping of the Emoticons based on the Emotions.}
\label{tab:emoticons_description}
\end{table}

\subsection{Emotion Recognition -- Classifier C}
\label{sec:er-C-C}
\begin{figure}
  \centering
  \includegraphics[width = 9.5cm]{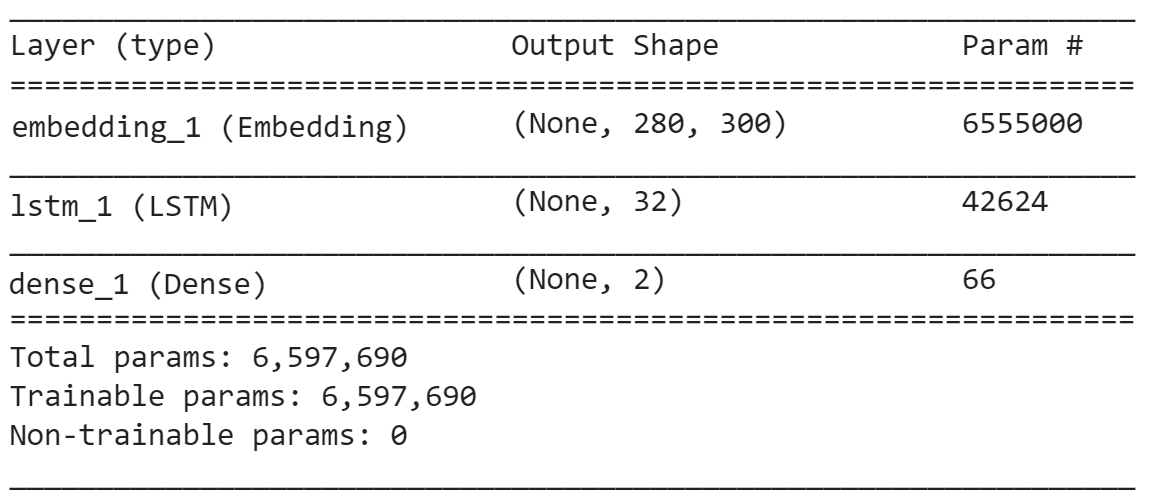}
  \caption{LSTM + GloVe Model Summary.}
  \label{model_summary_BC}
\end{figure}

The final classifier at the second stage is Classifier C, which classifies negative polarity tweets in negative emotions. As reported in Table \ref{table_emotion_dataset}, although there are four labels in negative emotion category, however, we drop the forth category -- disgust as it has very few instances and causes performance degradation for the dataset being imbalance. We performed experiments of remaining three labels on our five models. 
Table \ref{table_model_accuracy_sadness_anger_fear} summarises models' performance on 10\% test data. Once again, the classifier based on LSTM with pre-trained embedding Glove.twitter.27B.300d outperforms the other four models; therefore, we use it for classifying negative polarity tweets in negative emotions -- sadness, anger and fear. Figure \ref{model_summary_BC} shows the structure of model for classifier B and C.

The GloVe: Global Vector for Word Representation used in classifier B and C is a model for word representation trained on five corpora, a 2010 Wikipedia dump with 1 billion tokens; a 2014 Wikipedia dump with 1.6 billion tokens; Gigaword 5 which has 4.3 billion tokens; the combination Gigaword5 + Wikipedia2014, which has 6 billion tokens; and on 42 billion tokens of web data, from Common Crawl. The process of learning GloVe word embedding is explained in \cite{pennington2014glove}.

Similarly, FastText word embedding used in our Classifier A is an extension to word2vec model. FastText represents words as n-gram of characters. For example, to represent word \textit{computer} with n = 3, the FastText representation is \textit{<co, com, omp, mpu, put, ute, ter, er>}. A more detailed information on integration of general-purpose word embeddings like GloVe and FastText, and deep learning within a classification system can be found in \cite{Kastrati:2019ptr}.


\begin{figure}
  \centering
  \includegraphics[width = 16cm]{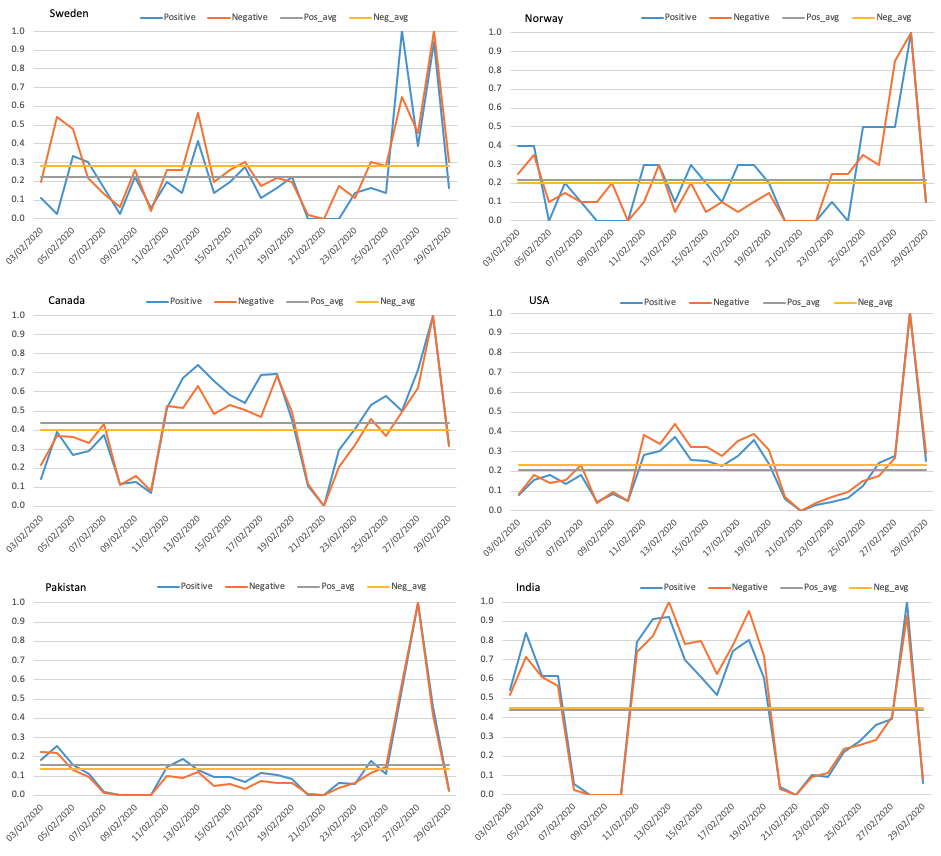}
  
 \caption{Side-by-Side Country-Wise Comparison of Sentiments Analysis on Trending Hashtag \# data for the Period Feb.3 to Feb.29, 2020. Positive and Negative Sentiment Graphs long with the Averaged Tweets' Polarity for Sweden \textit{(top--left)}, Norway \textit{(top--right)}, 
 Canada \textit{(middle--left)},
 USA \textit{(middle--right)}, 
 Pakistan \textit{(bottom--left)}, 
 and India \textit{(bottom--right)}.
 }
  \label{fig:Country-sentiment-Feb}
\end{figure}

\subsection{Validation Criteria}
\label{sec:validationCriteria}
The lack of ground truth i.e. labeled Tweets for queried test dataset for sentiment assessment concerning COVID-19, required the use of emoticons as a mechanism to validate the detected results into positive and negative polarities, as well as for emotions. We, therefore, propose the use of emoticons extracted from tweets to check whether a tweet's polarity and emotions reflect the sentiments depicted via emoticons the same or no. It may not be a perfect system, but a way to assess the accuracy of more than a million tweets via our proposed classifiers in a weakly supervised manner.  

The use of emoticons in sentiment analysis is not something new. In fact, there is an abundance of literature that supports the notion of utilizing emoticons in sentiment analysis \cite{debnath2020semantic, chang2020impact}. However, rather than using emoticons for sentiment detection, we use them for validating our model's performance. The emoticons were grouped into six categories, as described in Table \ref{table_emotion_dataset}. The type and description of emoticons used are depicted in Table \ref{tab:emoticons_description} for each group category.

We had a total number of 460,286 tweets from six selected countries in the English language. Out of these tweets, 443,670 tweets did not contain any emoticon, whereas 11,110 tweets used positive emoticon (joy, surprise), and 5,674 used negative emoticon (sad, disgust, anger, fear). The remaining tweets used a mix of emoticons like joy with disgust, anger with surprise, etc.; therefore, these usages of emoticons were considered sarcastic expressions of emotion, thus being excluded in the validation process.

We tested our Model \#2 presented in Table \ref{table_model_accuracy} (based on LSTM + FastText trained on Sentiment140) on the remaining 16,784 positive and negative tweets. We used these 16,784 tweets as test data to assess model accuracy. The emoticons were considered actual labels and the model predicted the labels on the tweet text. The model achieved an accuracy of 76\% and an F1 score of 78\%. This indicates that our model is reasonably consistent with the users' sentiments expressed in terms of emoticons.

The reason of good accuracy achieved in the validation phase is that our process of validation is indeed the same as the process used in preparing the Sentiment140 dataset – the dataset on which our model is based upon for sentiment polarity assessment.


\section{Results on Trending Hashtag \# Data}
\label{sec:results}
The proposed Model \#2, which achieved state-of-the-art polarity assessment accuracy on the Sentiment140 dataset, was used to detect polarity and emotions on the trending hashtag \# data. 

Figure \ref{fig:Country-sentiment-Feb} shows the side-by-side country-wise comparison of sentiment polarity detection for the initial period of four weeks. The sentiments are normalized to 0 - 1 as the sum of tweets per day/total number of tweets for a given country. As can be seen from the graphs illustrated in Figure \ref{fig:Country-sentiment-Feb}, there were only a few tweets concerning the coronavirus outbreak posted over almost all the month of February. There were also few days where no tweets have been posted, especially in Pakistan and India. It is interesting to note that the number of tweets is rapidly increased only in the last 2-3 days of February, and all six countries see this growing trend among Twitter users for sharing their attitudes, i.e., positive and negative about coronavirus.

The graphs between neighboring Sweden and Norway \textit{(top--row)} and that of Canada and USA \textit{(middle--row)} have a similar pattern of tweets' emotions, unlike Pakistan and India \textit{(bottom--row)}. In India, people's reaction seems quite strong, as evident from the average number of positive and negative posts (yellow and blue horizontal line). The reason could be the early outbreak of COVID-19 in India, i.e., $30^{th}$ of January 2020. A similar pattern was observed for Canada probably because they had their first positive case reported during the same time as well. 




\section{Discussion \& Analysis}
\label{sec:discussion}

\subsection{Polarity Assessment Analysis between Neighbouring Countries}
\begin{figure}
  \centering
  \includegraphics[width = 16cm]{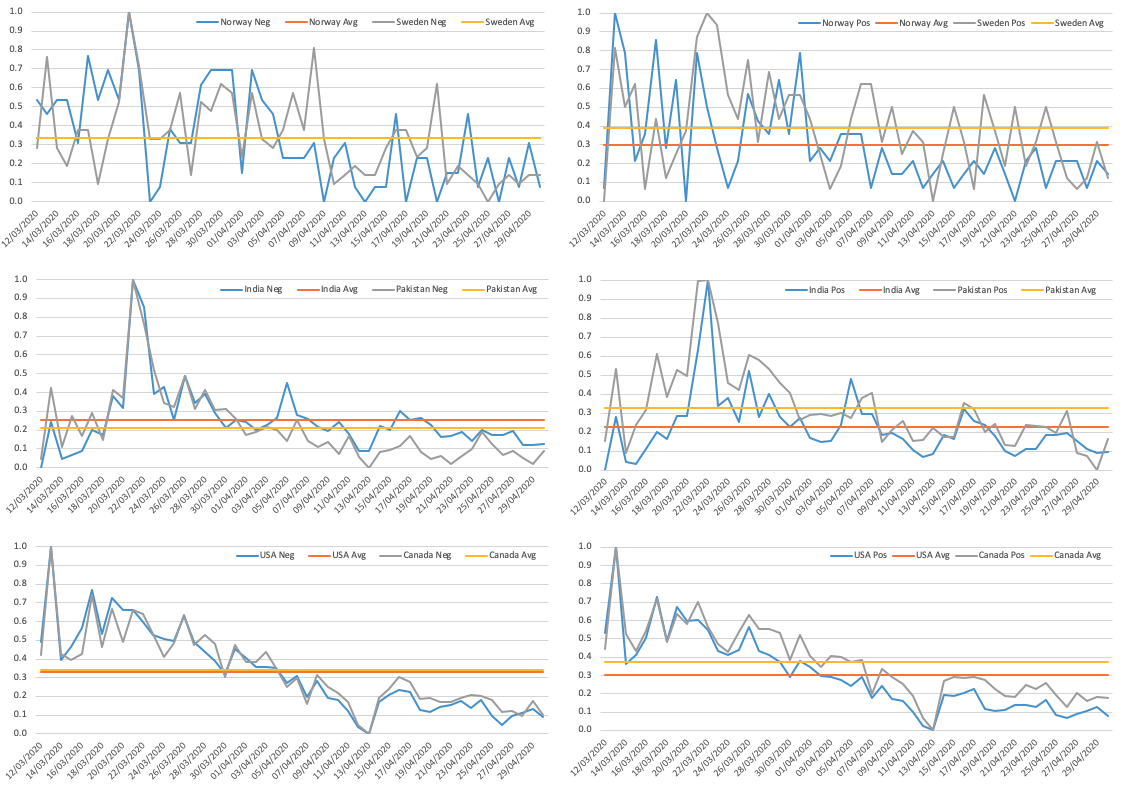}
  \caption{Side by side country-wise comparison of sentiments analysis:
   \textit{(top--left)} negative 
   polarity between NO -- SW, \textit{(top--right)} positive polarity between NO -- SW;
   \textit{(middle--left)} negative polarity between PK -- IN, \textit{(middle--right)} positive polarity between PK -- IN; \textit{(bottom--left)} negative polarity between US -- CA, \textit{(bottom--right)} positive polarity between US -- CA.}
  \label{fig:CountrySentiment}
\end{figure}

Figure \ref{fig:CountrySentiment} gives an overview of the side-by-side country-wise sentiment for both negative and positive polarity. The sentiments are normalized to 0 - 1. As can be seen in Figure \ref{fig:CountrySentiment} \textit{(top--left)}, the attitudes of Swedes over coronavirus outbreak has changed over time. The peak of negative comments expressed in twitter is registered on March 22. This was a day before the Prime Minister had a rare public appearance addressing the nation over the coronavirus outbreak. It is fascinating to note that on the day of Prime Minister's speech there exists an equal number of positive \textit{(top--right)} and negative \textit{(top--left)} sentiments, while a day after, the positive emotions dominated the tweets showing Swedes' trust in Government with respect to the outbreak.

\begin{figure}
  \centering
  \includegraphics[width = 16cm]{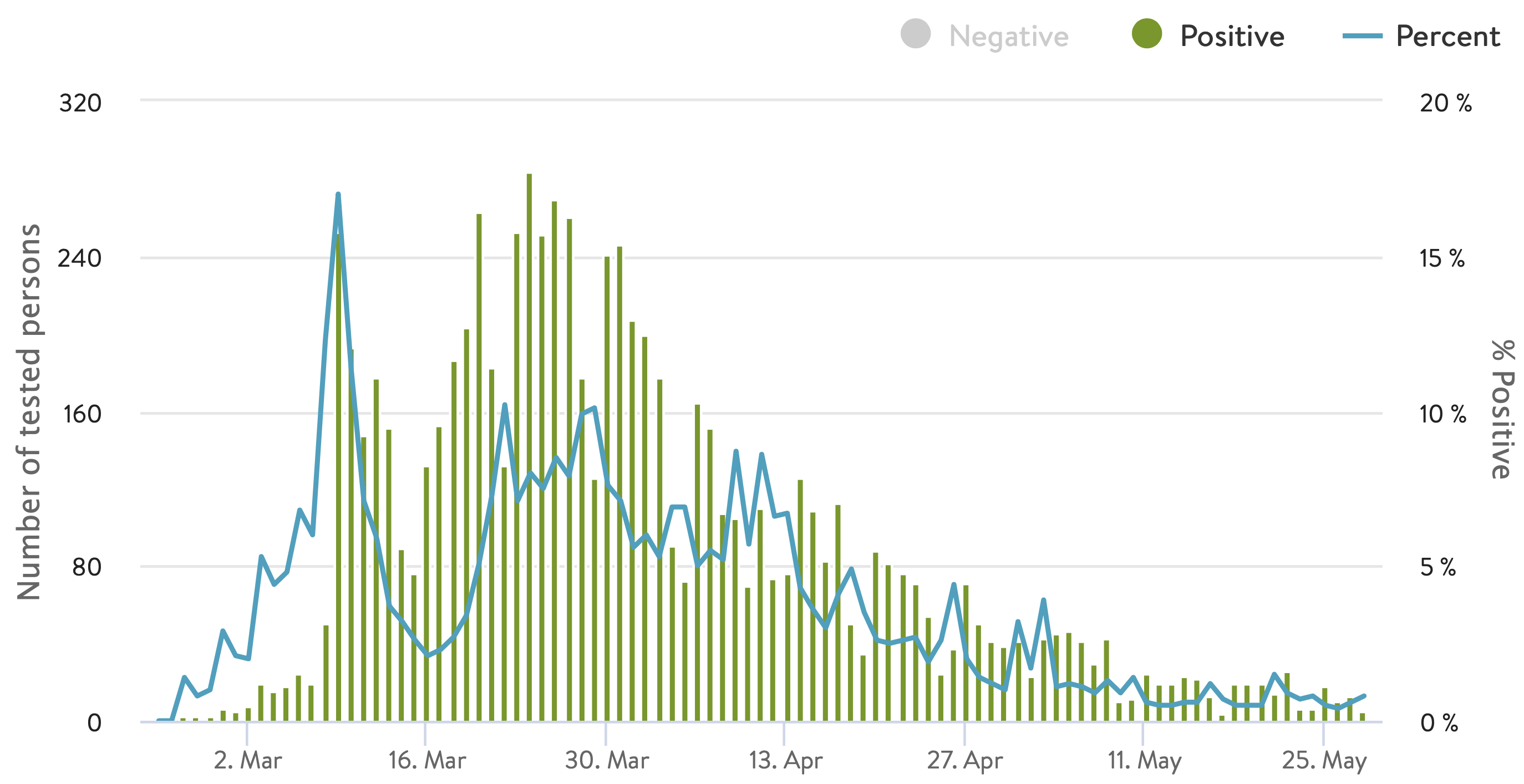}
  \caption{No. of positive cases reported in Norway between $24^{th}$ Feb. to $30^{th}$ May, 2020.}
  \label{fig:NorwayPositiveCase}
\end{figure}

There is an equal number of negative sentiments for both Norway and Sweden over the entire period, whereas the average polarity for positive sentiments is higher in the case of Sweden compared to Norway. A gradual decline in positive trends for Norway can be observed in \textit{(top--right)} plot in the figure. Till May $1^{st}$, 2020, the positive sentiments (blue line) for Norway were above the average (orange line), after which it started to decline. Figure \ref{fig:NorwayPositiveCase} shows the actual number of persons tested positive in Norway during the same period (data source\footnote{https://www.fhi.no/en/id/infectious-diseases/coronavirus/daily-reports/daily-reports-COVID19/}). The percentage of positive cases in the chart is based upon the total number of persons tested each day. The number of positive registered cases started increasing the second week of March 2020 till the first week of April, after which it dropped, which is in line with the sentiments expressed by the users which started to decline during the same week (Figure \ref{fig:CountrySentiment}, \textit{top--left and --right)}.

The trend between the positive and negative sentiments between Pakistan and India and that of the USA and Canada are very similar, as evident from the middle and bottom charts in Figure \ref{fig:CountrySentiment}. A closer look at the average sentiments between Pakistan and India reveals that the Indians expressed higher negative sentiments than Pakistanis \textit{(middle-left)}. Also, a significant number of positive posts appeared for Pakistan \textit{(middle--right)}, which showed that the people showed some trust in Government's decision. It partially is attributed to Pakistan's Prime Minister address to the nation on coronavirus on multiple occasions \textit{(March $17^{th}$ and March $22^{nd}$)} before the lockdown. It is worthy to note that the first case in India was reported on $30^{th}$ January 2020 and for Pakistan on $26^{th}$ February 2020, however, both countries went into the lockdown around the same time, i.e., $21^{st}$ of March for Pakistan and $24^{th}$ of March for India. Table \ref{tab:lockdown} shows when the first COVID-19 case was reported in the given country and the day it went into the lockdown.

\begin{table}[h]
\centering
\begin{tabular}{ |c|l|c|c| } 
 \hline
 Sr. \# & Country & First COVID-19 case & Lockdown Date \\
 \hline
 
 \hline
 1 & Pakistan & $26^{th}$ Feb. &	$21^{st}$ Mar.*\\ 
 \hline
 2 & India & $30^{th}$ Jan. &  $24^{th}$ Mar. \\ 
 \hline
 3 & Norway & $26^{th}$ Feb. & $12^{th}$ Mar. \\ 
 \hline
 4 & Sweden &\ $31^{st}$ Jan.	& No Lockdown \\ 
\hline
 5 & USA & $21^{st}$ Jan. & $19^{th}$ Mar.* \\
 \hline
 6 & Canada & $25^{th}$ Jan.	& $26^{th}$ Mar. \\
 \hline
  \end{tabular}
  
 \caption{Initial COVID-19 Case and the Lockdown Dates. *Lockdown Dates Varies for Different States/Province}
\label{tab:lockdown}
\end{table}


It is also worth mentioning here that at the beginning of April, the number of tweets declined, and so does the sentiments representation, which dropped below the average for all the countries except Sweden, where still a significant number of positive sentiments can be observed \textit{(top--right)}. Moreover, Pakistan had the least negative sentiments (i.e., avg = 0.201 - yellow line - \textit{(middle--left)}), whereas, Swedes were more positive (i.e., avg = 3.98 - yellow line - \textit{top--right)}). This could be attributed to the fact that most of the businesses run as usual in Sweden. In the case of Pakistan, the number of cases during the initial period was still low, as anticipated by the Government. Additionally, people did not observe the standard operating procedures enforced by the state much, despite the country was in lockdown. A similar trend was observed in India; however, the Government there had a much strict shutdown, though it came quite late since the first case was reported late January, which may have triggered more negative posts than positive.


\begin{figure}
  \centering
  \includegraphics[width = 16cm]{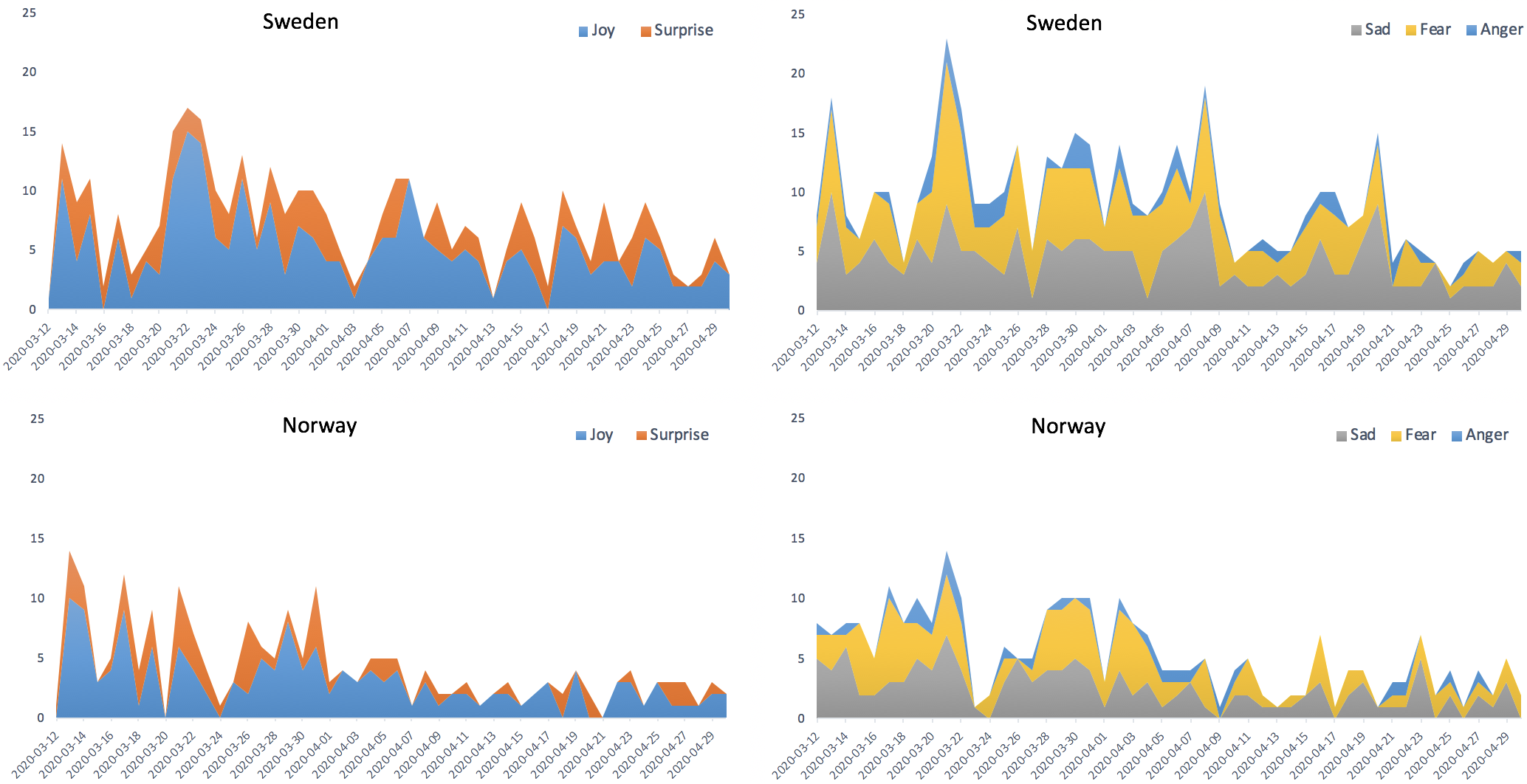}
  \caption{Side-by-Side Country-Wise Comparison of Emotions Between Sweden and Norway: \textit{(left-side)} +ve emotions, \textit{(right-side)} -ve emotions.}
  \label{fig:CountryEmotions}
\end{figure}

\subsection{Emotion Assessment Analysis Between Neighbouring Countries}

We observed that there was a visible difference between the sentiments expressed by the people of Norway and Sweden (Figure \ref{fig:CountrySentiment}). We further analyze these two countries in detail in our study of emotional behavior. The results are depicted in Figure \ref{fig:CountryEmotions}. Positive emotions are presented in the left figures and negative emotions on the right -- the graph shows which emotions are dominated over a period of time. The graph is scaled between 0 to 25 for better readability. It represents the accumulative emotions stacked on top of each other.   

As we can see from Figure \ref{fig:CountryEmotions} \textit{(top--left)}, in both countries, the joy dominates the positive tweets whereas sad and fear are the most commonly shared negative emotions, with anger being less shared. The pattern, in particular, for Norway is in line with the actual statistics for positive cases reported by the Norwegian Institute of Public Health (NIPH) (Figure \ref{fig:NorwayPositiveCase}).

\subsection{Correlation Analysis}
Additionally, we analyze the Pearson correlation between neighboring countries to see the sentiment polarity and emotion trend during the COVID-19 lockdown. As can be seen in Table \ref{tab:correlation_Sentiments}, there is a high correlation between USA and Canada (US-CA), and Pakistan and India (PK-IN), unlike between Norway and Sweden (NO-SW). The correlation between (NO-SW) is around 50\% for negative and 40\% for positive sentiments. This shows that the sentiments expressed in tweets on Twitter by the people of both countries were different during the same period. A possible reason for this is the different approach that these two countries have taken over the outbreak.

\begin{table}[h]
\centering
\begin{tabular}{ |c|l|c|c|c| } 
 \hline
 No.  & Sentiments & US-CA & PK-IN & NO-SW \\ 
  \hline

 \hline
 1 & Positive  & 0.967  & 0.816 & 0.402\\ 
 \hline
 2 & Negative & 0.971 & 0.860 & 0.517 \\ 
\hline
 
 \end{tabular}
 \caption{Correlation for Sentiment Polarity Between Neighbouring Countries.}
\label{tab:correlation_Sentiments}
\end{table}

Similar trend can be observed for emotions depicted in Table \ref{tab:correlation_emotions}. Pakistan and India have the highest correlation across all five emotions, followed by the USA and Canada. While Norway and Sweden have the least number of tweets sharing common polarity, as evident from the emotions "surprise" and "anger". A possible explanation for this is the response of people to respective Governments' decision on COVID-19, especially to the lockdown restrictions. There were few Swedes who felt surprised and angry as well, towards the Swedish Government's decision to not impose any lockdown measures and its choice to go for the herd immunity. For example, the tweet 
\begin{quote}
\textit{\lq\lq{}Tweet No 416: A mad experiment 10 million people \#Coronasverige \#COVID19 \#SWEDEN\rq\rq{} }
\end{quote}
expresses both feelings, surprise and anger, of the user on the decision of the Swedish Government. On the other side, users from Norway did not express any kind of these feelings as their Government followed the approach implied by most of the countries in the world by imposing lockdown measures from the very beginning of the outbreak. For instance, 

\begin{quote}
\textit{\lq\lq{}Tweet No 103: Norway closing borders, airports, harbours from Monday 16th 08:00. The Norwegian government taking Corona \#Covid\_19 seriously I wish us best hope survive\rq\rq{}}
\end{quote} 
shows people's faith in the Norwegian Government's decision.

\begin{table}[h]
\centering
\begin{tabular}{ |l|p{1.7cm}|c|c|c|c|c| } 
 \hline
 No  & Correlation b/w & Joy & Surprise & Sad & Fear & Anger\\ 
  \hline
  
 \hline
 1 & US-CA  & 0.795  & 0.740 & 0.877 & 0.718 & 0.673\\ 
 \hline
 2 & PK-IN  & 0.962  & 0.959 & 0.953 & 0.945 & 0.913\\ 
 \hline
 3 & NO-SW  & 0.229  & 0.161 & 0.343 & 0.375 & 0.190\\ 
 \hline
 
 \end{tabular}
 \caption{Correlation for Emotions Between Neighbouring Countries.}
\label{tab:correlation_emotions}
\end{table}

\subsection{Findings concerning RQ's}

Following the detected sentiment and emotions by the proposed model and the analysis of results presented in previous subsections, for \textit{(RQ1)}, it is safe to assume that NLP-based deep learning models can provide, if not enough, some cultural and emotional insight across cross-cultural trends. It is still difficult to say to what extent, as for non-native English speaking countries, the number of tweets was far less than those of the USA for any statistically significant observations. \textit{(RQ2)} Nevertheless, the general observations of users' concern and their response to respective Governments' decision on COVID-19 resonates with sentiments analyzed from the tweets. \textit{(RQ3)} It was observed that the there is a very high correlation between the sentiments expressed between the neighbouring countries within a region (Table \ref{tab:correlation_Sentiments} and \ref{tab:correlation_emotions}). For instance, Pakistan and India, similar to the USA and Canada, have similar polarity trends, unlike Norway and Sweden. \textit{(RQ4)} Both positive and negative emotions were equally observed concerning \#lockdown; however, in Pakistan, Norway, and Canada the average number of positive tweets was more than the negative ones (Figure \ref{fig:Country-sentiment-Feb} and \ref{fig:CountrySentiment}).


\section{Conclusion}
\label{sec:conclusion}
This paper aimed to find the correlation between sentiments and emotions of the people from within neighboring countries amidst coronavirus (COVID-19) outbreak from their tweets. Deep learning LSTM architecture utilizing pre-trained embedding models that achieved state-of-the-art accuracy on the Sentiment140 dataset and emotional tweet dataset are used for detecting both sentiment polarity and emotions from users' tweets on Twitter. Initial tweets right after the pandemic outbreak were extracted by tracking the trending hasthtags\# during February 2020. The study also utilized the publicly available Kaggle tweet dataset for March - April 2020. Tweets from six neighboring countries are analyzed, employing NLP-based sentiment analysis techniques. The paper also presents a unique way of validating the proposed model's performance via emoticons extracted from users' tweets. We further cross-checked the detected sentiment polarity and emotions via various published sources on the number of positive cases reported by respective health ministries and published statistics.   

Our findings showed a high correlation between tweets' polarity originating from the USA and Canada, and Pakistan and India. Whereas, despite many cultural similarities, the tweets posted following the corona outbreak between two Nordic countries, i.e., Sweden and Norway, showed quite the opposite polarity trend. Although joy and fear dominated between the two countries, the positive polarity dropped below the average for Norway much earlier than the Swedes. This may be due to the lockdown imposed in Norway for a good month and a half before the Government decided to ease the restrictions, whereas, Swedish Government went for the herd immunity, which was equally supported by the Swedes. Nevertheless, the average number of positive tweets was higher than the average number of negative tweets for Norway. The same trend was observed for Pakistan and Canada, where the positive tweets were more than the negative ones. We further observed that the number of negative and positive tweets started dropping below the average sentiments in the first and second week of April for all six countries. 

This study also suggests that NLP-based sentiment and emotion detection can not only help identify cross-cultural trends but is also plausible to link actual events to users' emotions expressed on social platforms with high certitude, and that despite socio-economic and cultural differences, there is a high correlation of sentiments expressed given a global crisis - such as in the case of coronavirus pandemic. Deep learning models on the other hand can further be enriched with semantically rich representations using ontology as presented in \cite{Kastrati:2015SITIS,kastrati2019impact} for effectively grasping one's opinion from tweets. Moreover, advanced seq2seq type language models as word embedding can be explored as a future work.

Till to date \textit{(i.e., the first week of May 2020)}, the pandemic is still rising in other parts of the world, including Brazil and Russia. It would be interesting to observe more extended patterns of tweets across more countries to detect and assert people's behavior dealing with such calamities. We hope and believe that this study will provide a new perspective to readers and the scientific community interested in exploring cultural similarities and differences from public opinions given a crisis, and that it could influence decision makers in transforming and developing efficient policies to better tackle the situation, safe-guarding people's interest and needs of the society. 

\bibliography{access.bib}{}
\bibliographystyle{IEEEtran}

\end{document}